\documentclass[11pt, a4paper]{article}

\usepackage[utf8]{inputenc}
\usepackage[T1]{fontenc}
\usepackage{amsmath, amssymb, amsfonts} 
\usepackage{graphicx}                    
\usepackage{geometry}                    
\usepackage{hyperref}                    
\usepackage{natbib}                      
\usepackage{float}                       
\usepackage{ulem}                        
\usepackage{booktabs}                    
\usepackage{array}
\usepackage{subfigure}
\usepackage{amsthm}                      

\title{\textbf{Social Norm Diffusion in Societies: Simple or Complex?}}
\author{
    \textbf{Claudia Lorenzetti$^{1,2}$, Aron Szekely$^{2,3,4}$, Daniele Vilone$^{2,5}$, Giulia Andrighetto$^{2,6,7}$}\\
    \textit{$^1$ CoMuNe Lab, Dept. of Physics and
Astronomy, University of Padua, Italy}\\
    \textit{$^2$ Institute of Cognitive Sciences and Technologies (ISTC), CNR, Rome, Italy}\\
    \textit{$^3$ Collegio Carlo Alberto, Turin 10122, Italy}\\
    \textit{$^4$ University College London, London WC1E 6BT, United Kingdom}\\
    \textit{$^5$ Grupo Interdisciplinar de Sistemas Complejos, Universidad Carlos III de Madrid, Spain}\\
    \textit{$^6$ Institute for Futures Studies, Stockholm, Sweden}\\
    \textit{$^7$ Institute for Analytical Sociology, Linköping University, Norrköping, Sweden}\\
    \texttt{claudia.lorenzetti2000@gmail.com}
}
\date{\today}

\begin{document}

\maketitle

\begin{abstract}
\noindent Social norms are a fundamental component of society, shaping how we communicate, cooperate, and avoid conflict. Despite their ubiquitous presence and profound influence on behaviour, questions remain about how they spread from person to person and whether they are adopted. Using an evolutionary game-theoretic model, we study whether simple or complex contagion processes characterise how social norms spread. Crucially, contagion type is endogenous and emerges from the model. We also vary cultural conformism, the cost of following a norm, and network topology to assess the generality of our results. We show that social norms in our model can spread through simple contagion when a new norm is low in cost and cultural conformism is low, while norms spread through complex contagion when the new norm is costly and cultural conformism is high.
\end{abstract}

\noindent\textbf{Keywords:} Social Norm Change, Tipping Point, Social Contagion.

\ 

\section{Introduction}
How do social norms, which shape how we talk, whether we help those in need, or protest against authoritarian governments, spread from person to person within societies? Under which conditions can population-level norms can emerge and spread starting from individuals interactions? Are simple contagion processes, where one independent exposure induces the adoption of a social norm, sufficient? Or are complex contagion processes needed, whereby independent reinforcement from multiple peers is required before adoption occurs? Here we develop a formal model of norm adoption to study this question in varied cultural environments, for norms that impose different costs on the individuals following them, and in multiple network structures. 

Social norms, behavioral rules that proscribe or prescribing actions and that are followed because of reciprocal expectations and, potentially, the threat of social punishment, are one of the most important mechanisms involved in the regulation of social behavior in social systems \cite{Bicchieri_2005} \cite{Cialdini1990} \cite{elster1989nuts} \cite{Gelfand2011} \cite{Szekely2021}. From how we dress, to how we greet, what is taboo or allowed, how we cooperate (or not), and even when we uphold "honour" and fight, social norms deeply influence our lives. As others have put it, "from forceps to grave, human life is wrapped in a tightly woven tapestry of rules, standards, and expectations that govern every aspect of social behavior" \cite{anderson2014behavioral}.

Reflecting this importance, a large body of research across the social, psychological, behavioral and computational sciences \cite{legros2020mapping} has studied, and advanced, our understanding of what social norms are, how they affect behavior, and how they are formed \cite{bicchieri2006grammar} \cite{Bicchieri2016} \cite{andrighetto2022research} \cite{przepiorka2020ruxolitinib} \cite{Szekely2021}. At the same time, our knowledge of the processes through which social norms spread and are transmitted from person to person remains surprisingly limited. 
To advance this state of affairs, we draw on a fundamental distinction that has been made in the diffusion of innovations literature: simple contagion \textit{à la} Granovetter \cite{granovetter1973strength} \cite{granovetter1974getting} \cite{granovetter1983strength} and complex contagion \cite{complexcontagion}. Simple contagion consists in a process in which a single exposure from one actor to another is sufficient for transfer or adoption to occur. Gossip, job advice, or contagious diseases are all phenomena that spread through simple contagion.

In this account, weak ties, that connect actors based on infrequent and superficial contacts and that typically link socially distant individuals, facilitate rapid and widespread diffusion. The strength of weak ties theory was famously proposed by Granovetter \cite{granovetter1973strength} and has also received empirical support  in various domains, including the diffusion of labour market opportunities \cite{granovetter1974getting} \cite{rajkumar2022evaluating}, learning in financial institutions \cite{uzzi2003relational}, and the apparel industry \cite{uzzi1996sources} to name a few.

Weak ties can facilitate the diffusion of information and the spread of contagious diseases, but not all diffusion processes can be adequately described within this framework. 

Hence, the need of the  definition of complex contagion \cite{complexcontagion} as a diffusive process that depends on social reinforcement from multiple ties, especially when we deal with the adoption of collective behaviors that entail substantial costs or risks. For an individual to adopt a given behavior, it may therefore be necessary for several members of their social environment to provide independent signals of approval, credibility, or legitimacy.

This distinction has important implications for the role of network topology in diffusion processes. The topology that favors the rapid dissemination of information may not be equally effective for the propagation of behaviors requiring social reinforcement. Simple contagions benefit from long-range ties connecting otherwise distant regions of a network, since a single exposure can be sufficient for transmission. Complex contagions, in contrast, require multiple independent sources of activation and therefore benefit from redundant connections and sufficiently wide bridges between groups \cite{complexcontagion}. Thus, while long ties can efficiently reduce path lengths and facilitate the spread of simple contagions, they may hinder complex contagions when they connect otherwise weakly coupled groups through only a small number of independent paths. More generally, the effectiveness of a given network topology cannot be considered independently of the underlying diffusion mechanism: the relevant network structure is determined jointly by the topology of interactions and the rule governing agents' adoption dynamics. 
In fact, social networks can exhibit markedly different structural properties. For instance, a society may be represented through an high level of clustering, as in the case of geographically localized communities or neighborhoods within large cities \cite{holland1983stochastic}. Alternatively, other settings, such as online follower networks, present highly heterogeneous connectivity, with a small number of highly connected hubs and a large number of nodes with comparatively few connections, as captured by scale‑free network models \cite{Baraba_si_1999}. Finally, in other contexts, such as friendship networks inside organizations, connections are predominantly local, yet a limited number of long‑range “shortcuts” substantially reducethe characteristic path length of the network, as captured by \textit{small-world models} \cite{watts1998collective}. These different topological structures can therefore lead to substantially different conditions for the emergence and diffusion of collective cultural changes.

In this framework, adoption dynamics are modulated by additional parameters. First, individuals vary in their susceptibility to social influence. We represent this with a degree of conformism, which captures how strongly people are influenced by others’ behavior. This parameter links the model to the tight–loose cultural distinction \cite{Gelfand2011}. Tight societies are characterized by strong, clearly defined norms and greater sanctions for violations. In contrast, loose societies are more tolerant of deviation, allowing greater freedom to adopt behaviors that depart from prevailing conventions. This tight–loose dichotomy has been investigated theoretically with models of the evolutionary dynamics of norms and their rates of change. In such models, conformism plays a central role as the parameter governing the tendency to follow established norms: higher conformism is associated with tighter societies, whereas lower conformism is linked to looser societies with greater flexibility to adopt alternative behavioral strategies \cite{de2018tipping}.

Second, the cost of complying with a norm affects its diffusion. Intuitively, norms related to fashion or greeting are easier to transmit because compliance is low-cost, whereas norms promoting protest or generosity diffuse more slowly because compliance is costly. These costs depend not only on the norm’s content but also on the social context. In relatively loose societies, new behaviors may spread even when they offer only modest gains in convenience or social payoff. In tighter societies, by contrast, adoption may require substantially greater benefits to offset the costs of deviating from established patterns of behavior.

For social norms, we lack a principled method to infer whether diffusion follows simple or complex contagion without assuming it a priori. We also lack a general account of how key features of norms and social contexts shape the macroscopic signatures of diffusion observed in aggregate data.

Inferring the nature of contagion for social norms is empirically difficult for at least three reasons.  First, normative influence is intertwined with conformity pressures and sanctions, making it hard to isolate the effect of multiple exposures from correlated adoption and homophily. Second, the material, psychological, and reputational costs and benefits of compliance vary widely across contexts, shifting the adoption threshold. Third, diffusion unfolds on heterogeneous networks whose clustering, path lengths, and degree distributions can accelerate or impede spread in ways that mimic simple or complex contagion.
 
Despite extensive work on contagion, network structure, and social norms, these dimensions have largely been studied separately. It remains unclear whether the distinction between simple and complex contagion is an intrinsic property of a behavior or an emergent consequence of interactions among behavioral incentives, social conformism, and network structure.

Taken together, these factors make it difficult to infer mechanisms from aggregate adoption trajectories alone and raise the risk that evidence for complex contagion is either under- or over-inferred.

This paper offers a framework for addressing these challenges. We develop a formal model of norm diffusion that does not specify \textit{ex ante} whether contagion is simple or complex. Instead, the adoption process is microfounded on three theoretically grounded dimensions that govern normative behavior: (i) topology, capturing the structure of social ties through which exposure accumulates; (ii) convenience, capturing the net costs and benefits of compliance; and (iii) conformism, capturing the strength of social pressure to align with others. We vary these dimensions systematically and observe the aggregate diffusion signatures that emerge.

First, we compare generative network models to identify the structural conditions under which the dynamics of interest arise. We assess whether the same behavioral model produces comparable outcomes across topologies. The dynamics are most clearly expressed under specific conditions in Watts–Strogatz networks, whereas other structures yield qualitatively different outcomes. This highlights that the effects of behavioral parameters must be interpreted relative to the structural conditions under which diffusion occurs.
Having identified the relevant network structure, we then examine how variations in its properties interact with norm convenience and individual conformism, allowing norm change to arise endogenously from their interplay.

This design enables post hoc inference about mechanism from aggregate data: if a real-world diffusion exhibits the constellation of signatures associated with a given region, we can classify its likely contagion as simple or complex type without prespecifying it in the model.

Our findings show that the same network structure can produce different diffusion dynamics depending on the population’s behavioral characteristics.

Thus, the interplay among norm convenience, agent conformism, and the network structure of their relationships yields nontrivial social dynamics, with the adoption or rejection of a new norm determined by the balance of these factors rather than imposed a priori by the model.

Moreover, network shortcuts have a dual effect: they can either accelerate diffusion or hinder it by facilitating competing behaviors. More generally, even a convenient norm may fail to spread when initial adopters are insufficiently connected, particularly in highly conformist environments. Exploring these dimensions in simulation, we find that diffusion can resemble either simple or complex contagion depending on parameter values, consistent with prior findings \cite{centola2015spontaneous}.
When costs for norm compliance are high and conformism is strong, reinforcement from multiple peers - namely complex contagion - becomes pivotal, producing slow early growth and sensitivity to clustered seeding. By contrast, when costs are low and conformism is weak, diffusion resembles simple contagion, with rapid growth and limited dependence on clustering.


Our contributions are threefold. First, we introduce a model that admits both simple- and complex-contagion-like outcomes without encoding them as assumptions, allowing macroscopic diffusion signatures to emerge endogenously. Second, we map how convenience, conformism, and topology jointly shape diffusion, linking substantive features of norms to observable aggregate behavior. Third, we show how variations in network structure induce distinct diffusion regimes, highlighting the interplay between local reinforcement and long-range social ties.

The remainder of the paper proceeds as follows. We present the model and define the three dimensions, then describe the simulation strategy. Next, we provide an analytical treatment of the model under stricter assumptions. Finally, we report results on how each dimension—and their interactions—shape diffusion patterns and the emergence of different contagion-like regimes.

\section{The model}

\subsection{Model motivation}

Does rationality drive all human actions, or do they depend also on the behaviors of the others? To what extent are individual decisions led by strategic considerations, and to what extent are they shaped by social influence?
Taking these questions into account, it seems appropriate to consider the interplay of these two \textit{driving forces}, rationality and conformism, when we want to study how social norms spread within a society, as in \cite{vilone2014social}, where both strategy and social imitative behavior play a key role in achieving consensus in a population. 
Hence, it is relevant how individuals from different populations react to the behavior of others. This leads to the concept of tight and loose societies \cite{Gelfand2011} \cite{centola2018experimental} \cite{centola2015spontaneous}: in particular, tight societies have a higher imitation behavior, while in loose societies the strategic component prevails.
\noindent At the same time, individuals do not decide to adopt a certain behavior if this is going to be too expensive for them. In fact, the personal cost associated with a behavior also affects the decision to follow it \cite{muthukrishna2020}.
Social influence takes place through a network, whose topology determines how individuals are exposed to one another, how social influence propagates through the population, and also the distribution of the social relationship among a certain social environment.

\noindent The goal is to integrate these concepts into the study of tipping points, shifting the focus from how quickly a new norm spreads to whether it will spread at all, and under which conditions.
To achieve that, the model in \cite{vilone2014social} will be adapted to the necessity of studying the tipping point: a more stochastic component to the strategic dynamic will be added, using an Evolutionary Game Theory approach very common in biology \cite{Moran_1958} and sociology. Furthermore, the new model will be applied to different network topologies to observe if the structure affects the outcome and to check whether the diffusion of a new collective behavior can be fostered by some specific social structures with respect to the others.
Through the study of these three main parameters of the model, we want to discover whether the diffusion patterns are reminiscent of a simple contagion or a complex contagion \cite{complexcontagion}. In fact, their interaction could reveal that the adoption requires simple exposure to the norm, that is typical of a simple contagion diffusion, or they could express the necessity of a repeated social reinforcement. 
We therefore do not impose simple or complex contagion as a microscopic rule. Instead, we ask whether the interaction between utility, conformity, and network structure can generate diffusion patterns resembling either mechanism.




\subsection{The three dimensions: conformism, convenience and topology}
\label{subsec:3d}

Before studying changes in social norms, it appears necessary to choose the right parameters to analyze the impact of their variations on the system, in such a way that underlines the relevant outcomes. 
Firstly, the whole model is based on the initial coexistence of two different norms: norm $A$, the newest and less diffused and norm $B$, the established one.
Evolutionary game theory provides this work with a solid framework to study which process drives either norms' diffusion or their absorption.
In our scope, all this dynamic is regulated by the systematic variation of three parameters. 
\noindent The first is the level of conformism $q$ of a society: the reaction of the system to the arrival of a new norm strictly depends on the tendency of each individual within that society to adapt to what their closest relationships do. Therefore, each agent can choose with a probability $q \in [0,1]$ whether to imitate without resistance what the others do or to behave in a more strategic way with a proportionality that depends on the level of conformism preassigned to their society. 
\noindent  The second parameter is convenience $\epsilon$, with $\epsilon \in [0,1]$. Since the new norm has not been widely adopted yet, agents would have no incentive to adopt it unless it offers an advantage over the existing norm. This is the reason why the collective behavior of the system clearly depends on how much the new norm is more convenient than the old one. 
\noindent Once we defined the rules that regulate how the single individual acts within a society, the last parameter regards the social structure, represented via networks. A network, in our case, is a graph $G(N,E)$, in which each node $N_i$ is an agent and each link $E_{ij}$ establishes a relationship between agent $i$ and agent $j$. 

Now that all the main parameters have been introduced, we can describe how the social dynamics evolves through time and how the systematic variation of all these three parameters let us study how the system reacts to norm changes.

\begin{figure}[t]
    \centering
    \includegraphics[width=\linewidth]{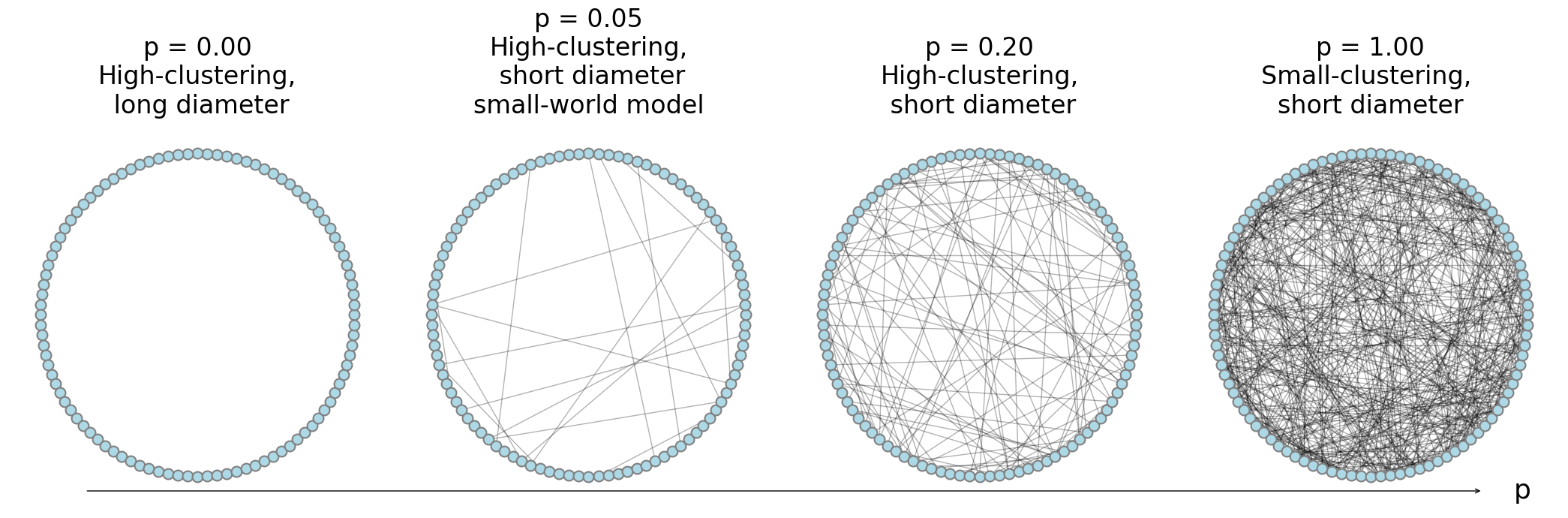}
    \caption{Different Watts-Strogatz networks as a function of the rewiring probability $p$. The small-world effect emerges for low values of $p$, between a fully regular network $p = 0$ and a fully random network $p = 1$. Graphs have been realized with $N = 100$ nodes and degree $k = 10$}
    \label{fig:WS}
\end{figure}

\subsection{The dynamic: an evolutionary game theory approach}
\label{subsec:networks}

Once we have described the main features of the model, the last step is to make it evolve over time. In this sense, Evolutionary Game Theory provides a powerful framework to study the conditions for new social norms to take over the existing order. The model extends the theoretical framework introduced in \cite{de2018tipping} adding more stochasticity and applying it to different network topologies to check whether the emergence of the new social norms works in the same way when the environment changes.
We employ a game-theoretical model based on a coordination game where agents improve their fitness adopting the same strategy of others: adopting a social norm is indeed an act of coordination on a specific context, which grants an advantage for all the individuals involved, while who does not coordinate with others loses in terms of fitness. Therefore, we describe the contagion of a new norm as a learning process where agents have to find that a different norm grants higher fitness if they coordinate on it instead of on the old one. On this basis, the payoff matrix of the elementary pair interaction has zero-entries outside the main diagonal, because a social norm holds only if individuals agree on it, while on the main diagonal there are the positive values representing the gain agents get in fitness when they coordinate on that specific norm. For simplicity, we consider the case of the competition between two norms, the old and the new, so that we can use a simple $2\times2$ matrix. 

\section{Methods}
\label{sec: methods}

\subsection{Topology}
\label{subsec:topology}

As discussed in Section~\ref{subsec:3d}, the topology of the interaction network plays a fundamental role in the  model. Since each graph $G(N,E)$ represents a different interaction structure, the underlying topology can substantially influence the resulting collective dynamics. Because this work focuses on the formation of norms and collective behavior in human communities, the interaction network should reproduce the main structural properties observed in empirical social systems.
Among the network models commonly adopted for social systems, two paradigmatic classes are small-world and scale-free networks. Small-world networks reproduce the coexistence of strong local clustering and short characteristic path lengths, whereas scale-free networks are characterized by a heterogeneous degree distribution with the presence of highly connected hubs. The latter have been widely used to describe online social networks and other systems in which interactions are strongly heterogeneous \cite{Baraba_si_1999} \cite{BARABASI200069} \cite{albert2002mean}. In this work, we mainly consider the Watts-Strogatz small-world model \cite{watts1998collective}, which provides a suitable framework for investigating the interplay between local interactions and long-range connections. For completeness, we also performed simulations on scale-free networks to assess the robustness of the observed dynamics with respect to the underlying topology.
The Watts-Strogatz model was introduced in \cite{watts1998collective} to interpolate between regular lattices and random graphs. Its main feature is the simultaneous presence of a high clustering coefficient and a short characteristic path length, two properties that are commonly observed in real social networks~\cite{caldarelli2012networks}. These characteristics make the model well suited for investigating how local interactions give rise to global patterns of collective behavior.

The network is generated by starting from a one-dimensional ring lattice with $N$ vertices, where each vertex is connected to its $k$ nearest neighbors. Each edge is then rewired independently with probability $p$, preserving the total number of edges while introducing long-range connections. The rewiring probability $p$ acts as a control parameter that continuously interpolates through different topologies (Figure \ref{fig:WS}): at the extremes we have regular lattice when $p=0$ and a fully random network with $p=1$, allowing the exploration of the intermediate regime $\frac{1}{N}\lesssim p\lesssim0.1$ in which the features of the small-world topology emerge \cite{watts1998collective},\cite{strogatz2001exploring}.

\subsection{Evolution dynamic}
\label{subsec:evdynamic}

As previously established, the whole dynamic happens on a graph $G(N,E)$ in which each node is an agent and the edges represent the social connections between them. 

\noindent At the very beginning of the simulation, the majority of the nodes are assigned to follow the old social norm $B$, while a smaller part is given to follow the new norm $A$.  This small cluster of new adopters is the seed of the new norm and its size is a crucial parameter to determine whether the new norm will be able to spread or not. The initial fraction of agents that follow the new norm $A$ is called $\rho_{A,0}$, while the fraction of agents that follow the old norm $B$ is $\rho_{B,0} = 1 - \rho_{A,0}$.

\noindent At each turn of the game, a random agent is selected: she can choose to act whether in an imitative way with a probability $P_{VM} = q$, following a classic voter model , or in a strategic way with a probability $P_{FR} = 1-q$. 

\noindent If she chooses to behave in a more strategic way, she will get a payoff $\pi_i$ according to the following table: 

\begin{center}
$\pi$\ \ \ = \ \ \ 
\begin{tabular}{|c|c|c|}
\hline
\mbox{ } & {\bf A} & {\bf B} 
 \\ \hline
{\bf A} & $1 + \epsilon,\, 1+ \epsilon$ & $0,0$
 \\ \hline
{\bf B} & $0,0$ & $1,1$ \\
\hline 
\end{tabular} \ . 
\end{center}

\ 

\noindent This payoff matrix is a clear representation of a coordination game in which the personal interest is in any case subjected to the will to cooperate to get a higher reward. 

\noindent The parameter $\epsilon \in [0,1]$ represents the relative convenience associated with adopting the new norm $A$ rather than maintaining the existing norm $B$. Naturally, such convenience is significant only to the extent that the agents interacting with the selected individual have already adopted the new norm. After each interaction, the agent updates her strategy according to the outcome of the dynamics, which may or may not result in a change of strategy.


    

\noindent Time in this model is measured in units of time $udt$ corresponding to Monte Carlo steps: this implies that one unit of time represents a round in which all $N$ agents have played one time on average. The simulation continues until a consensus is reached, meaning that a single norm prevails throughout the social system, or a maximum number of $udt$ has elapsed.

\subsection{Analytical approach}
\label{subsec: analytical approach}

In this section we aim to give a mathematical formalization to the model described above. The aim of this section is to write a master equation that can qualitatively describe the results.
\noindent We build an equation under mean field assumptions. Although it is an approximation since in our simulations the topology is more complex and the composition of the population is different with respect to an all-to-all dynamic, these results can still help to give a general interpretation to this kind of dynamic. 

Considering $n_A$ as the fraction of nodes $N$ that chose to follow the new norm $A$ we want to write the master equation as the time evolution of the probability of having $n_A$ agents at the time $t$, $P_{n_A}(t)$.

\noindent Hence, the discrete master equation has the following structure:

\begin{equation}
\begin{aligned}
  \partial_t P_{n_A}(t)
  &= T^-_{n_A + 1}\,P_{n_A+1}(t)+\\
  &\quad T^+_{n_A - 1}\,P_{n_A-1}(t)-\bigl[T^+_{n_A}+T^-_{n_A}\bigr]P_{n_A}(t)
\end{aligned}
  \label{eq: me}
\end{equation}

\noindent with $T^{\pm}$ being the transition rates.
\noindent In our case, an increase in the population $n_A$ may occur either due to the Voter Model with probability $q$, or due to the Fermi Rule with probability $1 - q$. Under a mean-field approximation, transition probabilities are obtained by combining the voter and Fermi updating rules. The overall transition rates are written as:

\begin{equation}
  T^+(n_A) = qT^+_{VM}(n_A) + (1 - q)T^+_{FR}(n_A)
  \label{eq:Tpiù}
\end{equation}
\begin{equation}
  T^-(n_A) = qT^-_{VM}(n_A) + (1 - q)T^-_{FR}(n_A)
  \label{eq:Tmeno}
\end{equation}

\subsection{Master equation}
\label{subsec: me_model}
After evaluating the transition probabilities, we perform a change of variable $\rho = \frac{n_A}{N}$ (that is just the fraction of agents adopting the $A$ strategy at time $t$) and we take the thermodynamic limit $N \to +\infty$. Therefore, in the continuous limit our master equation assumes this form (see in the Appendix for details):

\begin{equation}
\begin{aligned}
  \partial_tP(\rho,t) = \rho(1-\rho)\bigg[\bigg(\frac{1+q}{2}\bigg)\partial^2_\rho P(\rho,t) + \\
  (1-q)\tanh\!\left(\tfrac{\beta \Delta\pi}{2}\right)\partial_\rho P(\rho,t)\bigg] \ ,
\end{aligned}
  \label{eq:final}
\end{equation}

\noindent where $\Delta\pi$ accounts for the average difference between the two strategies that depends on the value of $\rho$.

Eq. \eqref{eq:final} naturally separates the contributions of the two updating mechanisms. The voter dynamics contributes a purely diffusive term, reflecting unbiased social imitation. By contrast, the strategic Fermi rule generates both an additional diffusive component and a drift term proportional to the payoff difference. The latter introduces a systematic bias towards strategies with higher expected payoff, thus competing with the stochastic spreading induced by social imitation. Of course, as it is easy to verify, if $\epsilon\rightarrow0$, only the diffusion contribution remains.
\noindent By the way, this mean-field assumption is useful to give an idea of the general dynamic of the model. It describes well how conformism and strategy are interconnected and their way of spreading the norm in the network.
However, such approximations like random assignments of the norms and the mean over the ensemble, require us to consider this as a useful tool, but a first approximation approach.
\noindent Since the system described by this approach is mathematically relevant but lacks sociological significance, it is necessary to consider more complex structures. The introduction of topological noise, together with variations in the initial conditions, makes the model too intricate to be treated with a purely analytical approach. Therefore, in order to support our argument, the following chapter presents a series of simulations that account for more complex topologies. 

\section{Results}
\label{sec: results}

We consider as the initial state a configuration in which the established norm \(B\) is adopted by all agents, with the exception of a small, compact cluster of agents who have adopted the alternative norm \(A\).

The size of this initial cluster is set to \(\rho_0 = 10\%\) of the total population, where \(N\) denotes the total number of agents. We then systematically explore the parameter space along the three dimensions described in \ref{subsec:3d}.

First, we investigate the role of the topology of the interaction network \ref{subsec: top affection}, showing that different network structures can give rise to qualitatively different outcomes. Then, in \ref{subsec: conf and conv}, we investigate the interplay between convenience and conformism, showing that their variation gives rise to a transition region in the parameter space. We then focus on the role of conformism in determining the dynamics around the tipping point. In \ref{subsec: conv time}, we analyze how different levels of conformism affect the convergence dynamics and the time required for the system to reach a stable configuration.

Finally, in \ref{subsec: complex contagion}, we analyze how variations in the structural properties of the network affect the diffusion dynamics. In particular, we show how different simple and complex contagion regimes can emerge by analyzing the convergence time as a function of the model's rewiring probability.

\subsection{Effects of topology on norm change}
\label{subsec: top affection}

\begin{figure}[H]
  \centering

  \begin{subfigure}
    \centering
    \includegraphics[width=9.6cm]{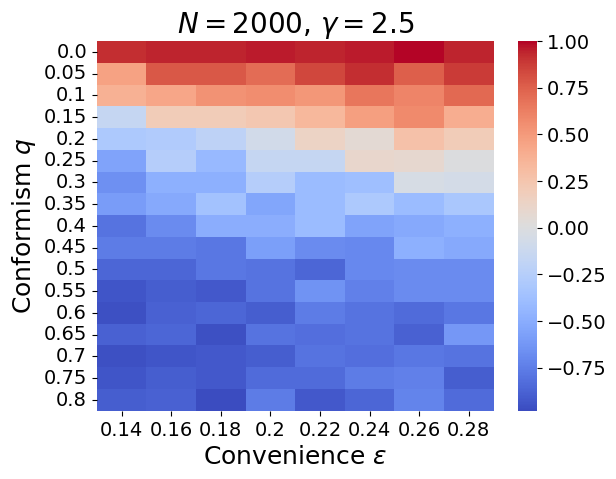}
  \end{subfigure}

  \par\medskip

  \begin{subfigure}
    \centering
    \includegraphics[width=9.6cm]{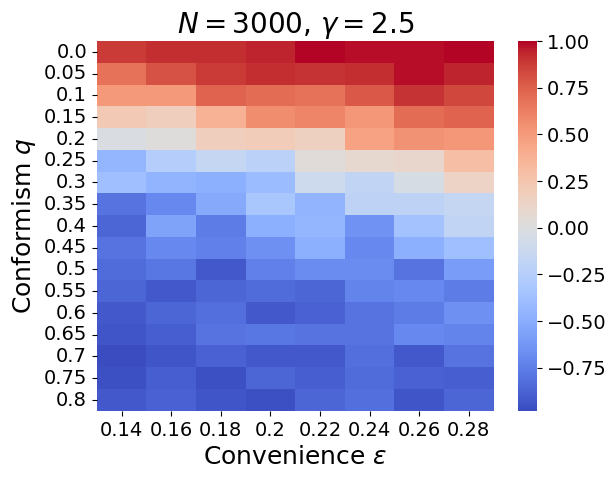}
  \end{subfigure}

  \caption{Heatmaps showing the fraction out of 100 repetitions in which the network converges either to the new norm $A$ or to the old norm $B$. The color scale goes from blue, which means each repetition converges to the old norm, to red, which means each repetition converges to the new norm $A$. The simulations are run both on a network of $N=2000$ agents (top) and on a network of $N=3000$ agents (bottom).}
  \label{fig:scalefree}
\end{figure}

In \ref{subsec:topology} two different kinds of networks have been shown: the Watts-Strogatz and the scale free network. The same simulation is run on both topologies, showing different results. In particular, the two network structures give rise to qualitatively different heatmaps in the parameter space. While the Watts-Strogatz network shows a transition region between the persistence of the old norm $B$ and the emergence of the new norm $A$ (Figure \ref{fig:WS}), the scale-free network displays a different pattern and no comparable transition region can be identified (Figure \ref{fig:scalefree}).  This indicates that topology is not merely a background for the dynamics, but represents a relevant feature of the model, affecting whether the transition dynamics can emerge.

Since the transition is observed under the specific conditions of the Watts-Strogatz network, we focus on this topology to systematically investigate how the variation of conformism and convenience shapes the transition between the two norms.

\subsection{Systematic variation of conformism and convenience leads to a tipping point}
\label{subsec: conf and conv}

We test the model on a Watts-Strogatz network, built as a small-world model with $N$ as the number of agents; each agent is connected to $k = 10$ neighbors and the rewiring probability is set to $p = 0.05$. 

\begin{figure}[H]
  \centering

  \begin{subfigure}
    \centering
    \includegraphics[width=9.6cm]{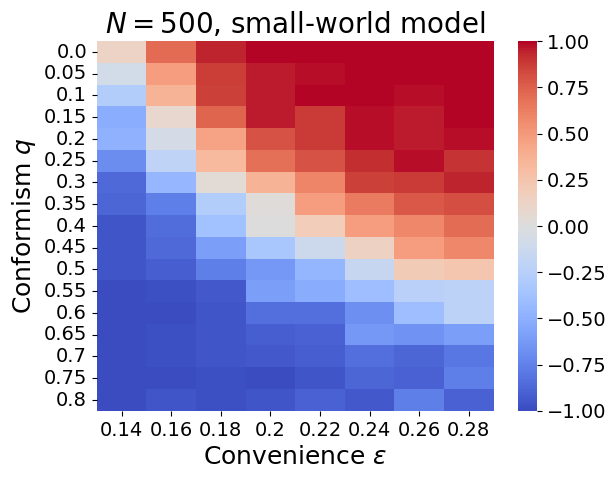}
  \end{subfigure}

  \par\medskip

  \begin{subfigure}
    \centering
    \includegraphics[width=9.6cm]{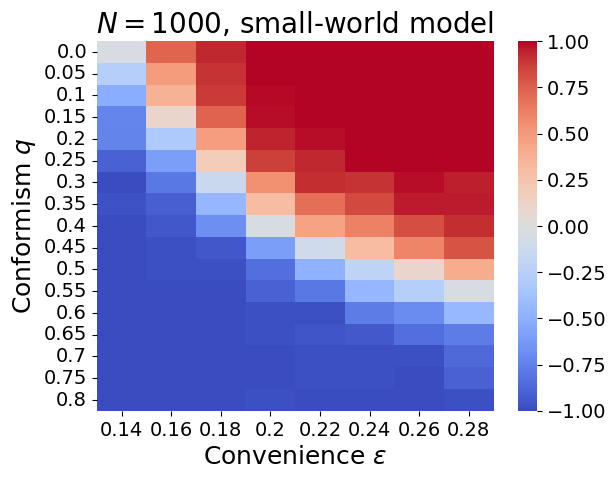}
  \end{subfigure}

  \par\medskip

  \begin{subfigure}
    \centering
    \includegraphics[width=9.6cm]{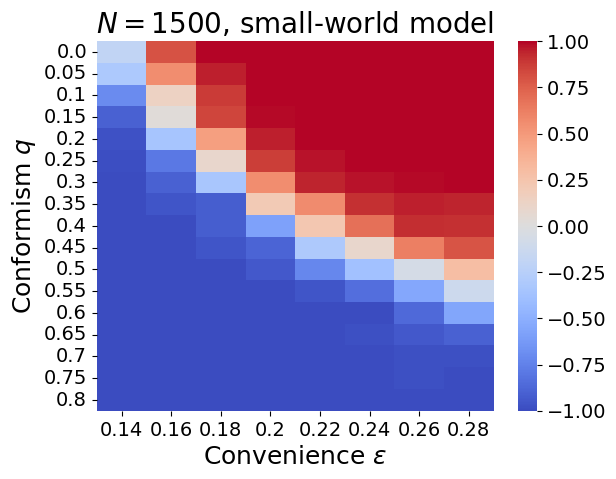}
  \end{subfigure}

  \caption{Heatmaps of the fraction of repetitions in which the network converges either to the old norm $B$ or to the new norm $A$, for different population sizes. From top to bottom: $N=500$, $N=1000$, and $N=1500$. The x-axis represents the convenience $\epsilon$, while the y-axis represents the conformism $q$. Each point is the result of $100$ independent simulations. Note how the lighter area becomes shallower as $N$ increases.}
  \label{fig:WS_heatmaps}
\end{figure}

\noindent To spot a tipping point, we varied the conformism $q$ and the convenience $\epsilon$ keeping in mind that, in order to have non trivial results, the payoff $\pi_{A,A}$ cannot be too much higher than $\pi_{B,B}$. Once the range of the parameters is set, the simulations are run for $100$ independent repetitions. The results are shown in Figure~\ref{fig:WS_heatmaps}. The color scale represents the average fraction of simulations that reached consensus, red scale if the consensus is on the new norm $A$, blue scale if the consensus is on the old norm $B$. The simulations have been carried out for different population sizes, for $N = {500,1000,1500}$, to test how the model behaves as the system size increases.

\noindent The plots \ref{fig:WS_heatmaps} show a clear transition from the blue area, the region in which the old norm $B$ prevails, to the red area, in which the new norm $A$ is able to take over the whole system. It is relevant to notice that this transition becomes sharper as the size of the system increases, suggesting that, hypothetically, for $N \to +\infty$ the transition would become more similar to a curve in the $\epsilon-q$ space.

\noindent This transition clearly shows that, in order for the new norm to prevail despite the value of conformism within that specific society, the convenience $\epsilon$ must be suitable for the agents. In Figure \ref{fig:WS_1000}, for example, the red area becomes wider as the convenience $\epsilon$ increases. For low values like $\epsilon = 0.14$ the new norm does not spread through the population because changing it is expensive for an agent and, if the convenience is not high enough, the gain obtained is not sufficient.

\noindent Overall, Figure \ref{fig:WS_1000} illustrates that, at low levels of conformism $q$, the emergence of the new norm becomes predominant even when the associated gains are relatively small. In general, a strategic approach tends to favor the emergence of the new norm $A$, which is more convenient. The results clearly indicate the presence of a transition area between the two different norms.

\subsection{Convergence time}
\label{subsec: conv time}
\noindent After having analyzed the results to explore the possibility of having a turning curve in the parameter space, we focus on the convergence time, i.e., the time needed for the new norm to diffuse. 
\noindent In particular we are interested in what happens around the critical point both in systems characterized by high levels of conformism and in networks with low values of $q$, in which the strategic component prevails.

\noindent The hypothesis we want to test is that, in loose societies, where the strategic component is more relevant, convergence is reached more frequently but through a more gradual process, characterized by a progressive increase in the fraction of agents adopting the new norm.. On the other hand, in tight societies, where people tend to be more conformist, convergence to the new norm is less likely to happen, but the time needed to reach consensus in that case is lower, once the critical mass has been reached.

\noindent To check this hypothesis we conducted simulations only on a Watts-Strogatz network, since it is the only one that offers us a wider range of cases. We chose the same parameters as before, so $p = 0.05$, $k = 10$ and, in this case $N = 1000$. This time the convenience is fixed at  $\epsilon = 0.2$ and both the conformism $q$ and the initial cluster size $\rho_{0,A}$ are systematically varied. We run the simulations as before, for $100$ independent simulations for each point in the parameters space.

\noindent Once this heatmap has been obtained, we check the red area of the map and we chose for each value of conformism $q$ the minimum value of the initial cluster size $\rho_{0,A}$ that was able to reach consensus to the new norm in at least $70\%$ of the repetitions: this is how we selected our \textit{critical mass}.
\noindent Once we chose the critical mass $\rho^*_{A}$, we run the simulations again with $\rho_{0,A} = 10\%$ and registered the time needed to reach the critical mass $\tau^*$ and the time needed to reach consensus once the critical mass has been reached $\tau_c$.

\noindent As clearly shown above, the time needed to reach the critical mass $\tau^*$ grows as the conformism $q$ increases. This is something we expect since, the more a society is conformist, the more it will be difficult for a new norm to spread, especially if it is stood by just a small component of the population. 
\noindent On the other hand, the time needed to reach consensus $\tau_c$ once the critical mass has been reached is smaller for conformist societies. It seems that a conformist system, a tight society, struggles to change its norms, but, once a norm has been adopted by a good part of the population, the rest of the agents will follow it very fast (Figure \ref{fig:020_045}).
\noindent More in general, imitation requires less effort than acting with a strategy and this could be another reason why, once a consistent group has been convinced, the others follow them easily in tight societies.

\noindent For loose societies instead, the time needed to reach the critical mass $\tau^*$ is lower and, because of that, it happens more often. However, once the critical mass has been reached, the system needs more time to reach the consensus among the population (Figure \ref{fig:020_02}). This is, in a certain way, counterintuitive, but it can be explained with the fact that, in a loose society, the strategic component prevails and individuals tend to act more for their own interest rather than for the wellness of the group. This leads to a more gradual diffusion of the new norm, with agents that change their mind and then change it back again, until the whole system is convinced that the new norm is the best option.

\begin{figure}[H]
  \centering

  \begin{subfigure}
    \centering
    \includegraphics[width=11.6cm]{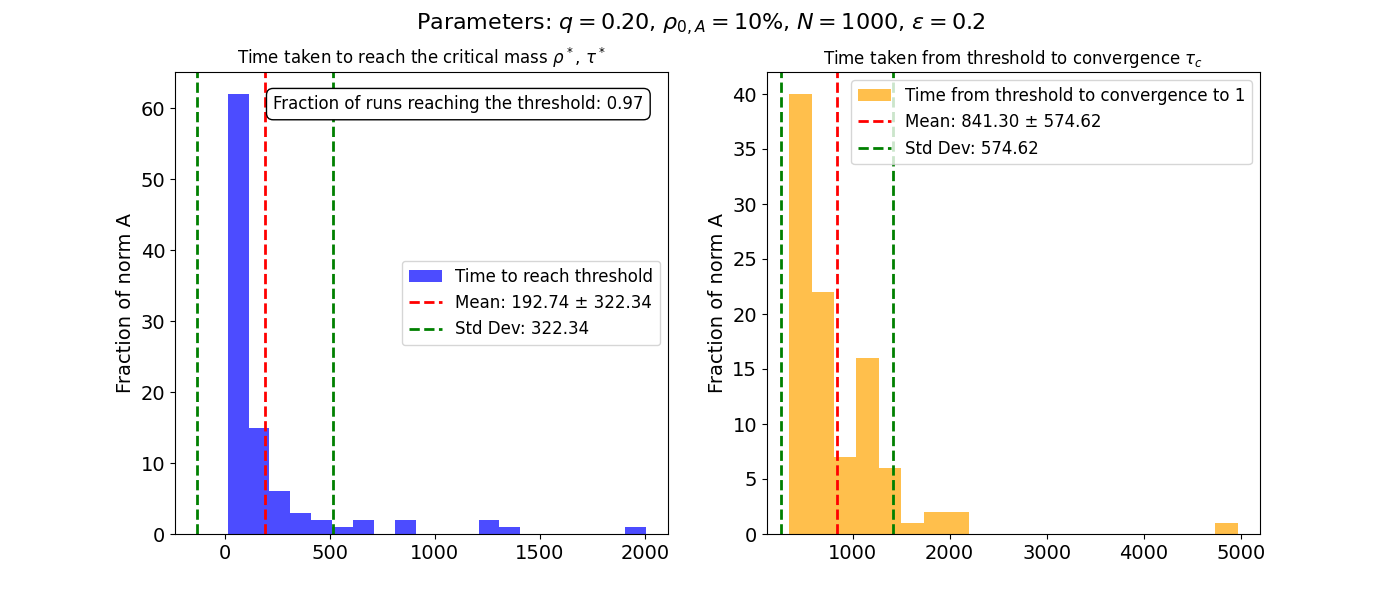}
  \end{subfigure}

  \par\medskip

  \begin{subfigure}
    \centering
    \includegraphics[width=11.6cm]{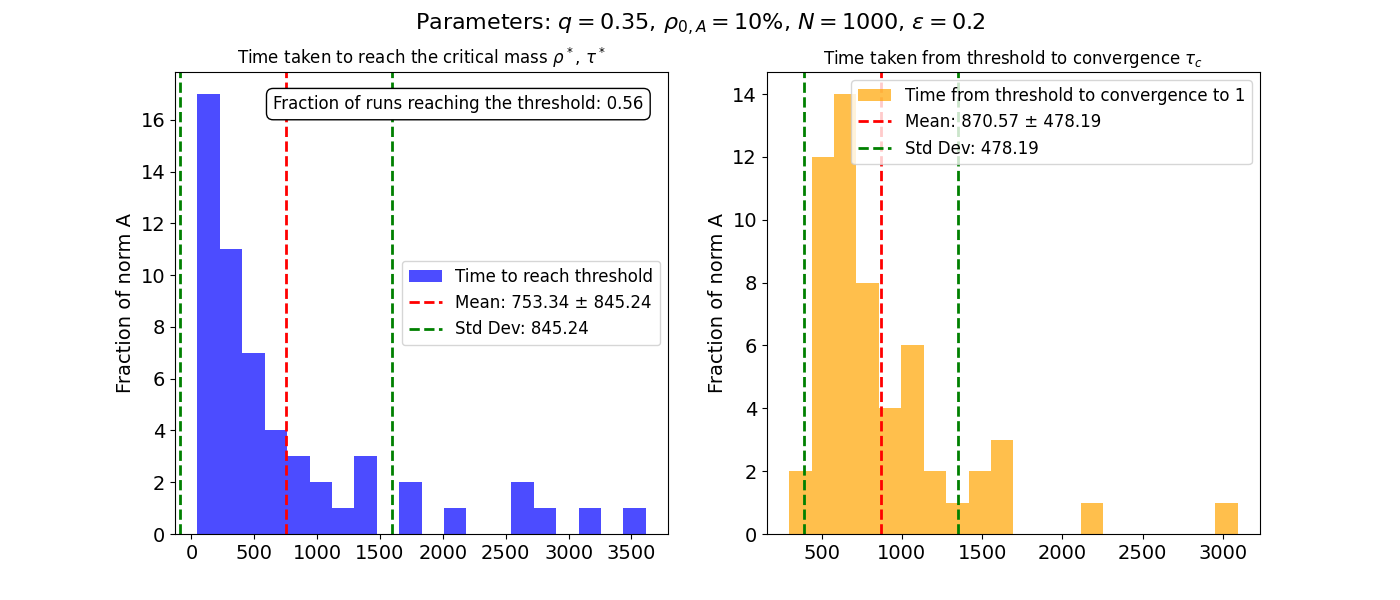}
  \end{subfigure}

  \par\medskip

  \begin{subfigure}
    \centering
    \includegraphics[width=11.6cm]{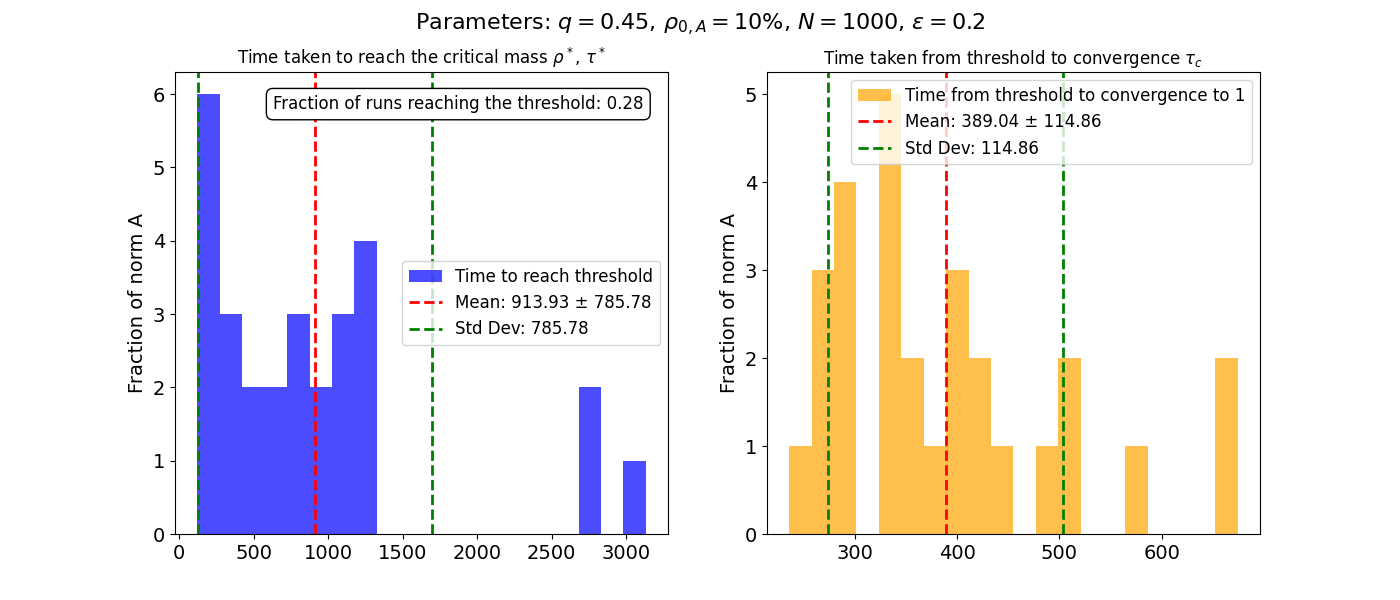}
  \end{subfigure}

  \caption{Histograms of the time needed to reach the critical mass $\tau^*$ (\textit{on the left}) and the time needed to reach consensus $\tau_c$ (\textit{on the right}). From top to bottom:  $q=0.20$, $q=0.35$, and $q=0.45$.}
  \label{fig:hist_conv}
\end{figure}

\begin{figure}[t]
  \centering
  \includegraphics[width=\textwidth]{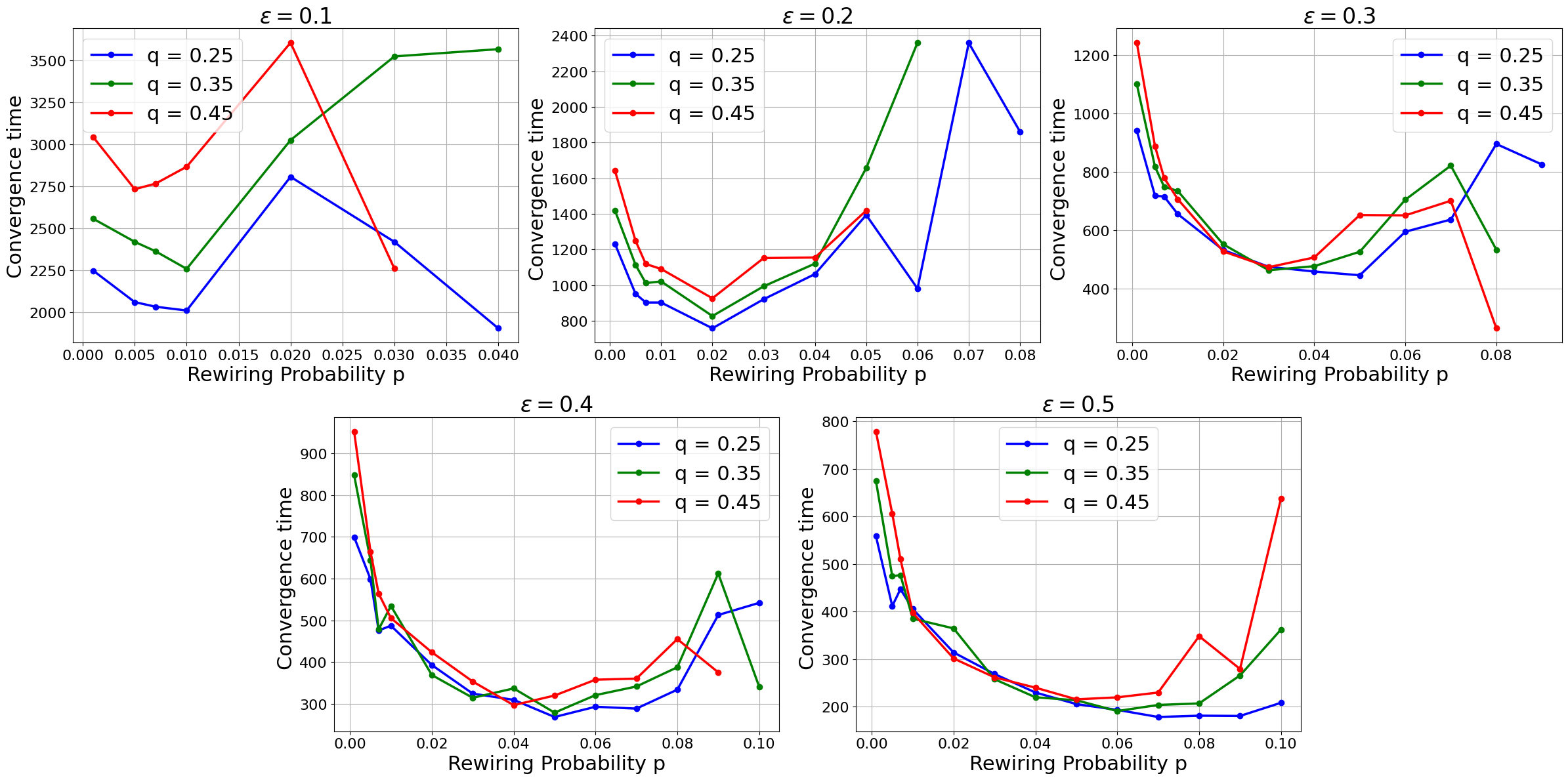}
  \caption{Convergence time needed for the new norm $A$ to take over as a function of the rewiring probability $p$ in a Watts-Strogatz network. The results are obtained for different value of the convenience $\epsilon$ and for three different levels of conformism $q$: below the transition area, within the transition area and over the transition area. The plots shows the different diffusive behavior of the model depending on the parameters. The second and the third plot, in particular, show a trend that can be associated with a complex contagion regime.}
  \label{fig:complex}
\end{figure}

\subsection{The rewiring probability variation and the emergence of complex contagion}
\label{subsec: complex contagion}

In \ref{subsec: top affection}, we highlight how differences in the social environment can lead to different outcomes. By simulating the same model on both a Watts–Strogatz and a scale-free network, we observe that the emergence of a transition phase is primarily associated with the network structure that more closely resembles real-world social systems: the \textit{small-world model}. 

\noindent Rather than directly studying the contagion process and subsequently assessing whether the resulting dynamics are consistent with the theoretical predictions, we adopt a more systematic approach based on varying the rewiring probability $p$. Specifically, we analyze the speed of convergence toward the new norm as a function of the topological parameter $p$, considering different values of conformism $q$ and convenience $\epsilon$. By examining how the convergence dynamics change across these parameter regimes, we can infer the influence of the underlying network topology and identify whether the resulting diffusion mechanism exhibits the characteristics associated with simple or complex contagion.
Hence, we systematically and significantly changed the value of the rewiring probability $p$, which refers to the probability of cutting an edge between two nodes and rewiring it, by linking one of the two, to another randomly chosen node. If $ 0.01 \leq p \leq 0.05 $ the network lives in the \textit{small-world model} regime, but if the $p$ changes value the topology can vary from a very random network to an ordered ring.
Figure \ref{fig:complex} shows the convergence time to the new norm for different values of conformism as the rewiring probability increases. From pictures (b), (c), (d) and (e) the convergence time seems to progressively become smaller until a minimum value is reached and, after that, it increases again. This behavior is similar to the one described in \cite{complexcontagion}, which shows that when the time of the diffusion does not decrease as the number of links is increased, the contagion becomes complex. In this case, the behavior of the curve resembles the one of a complex contagion. Different values of conformism produce similar, but offset curves. The level of convenience associated with the adoption of the new norm, denoted by $\epsilon$, has a significant impact on the results. As the value of this parameter increases, the behavior of the curve becomes progressively more linear, ultimately approximating the dynamics of a simple contagion process.

\section{Discussion}

We ask whether norm-change dynamics emerge in our model and which of the three core variables—network topology, conformism, and convenience—are determinant or redundant. We find that all three dimensions play essential roles: norm diffusion does not reduce to any single factor but emerges from their combined effects. Topology conditions the emergence of the dynamics: not all social structures yield the same outcomes. Once the most suitable composition has been found, varying conformism and the convenience of the norm reveals a tipping point: their joint variation leads either to system-wide conversion or to re-establishment of the status quo. Within the transition region, distinct diffusion dynamics emerge as social ties vary. Thus, the emerge of a new norm is a system property that arises from interactions among these dimensions rather than from any single variable.

The dynamics of interest emerge most clearly on Watts–Strogatz networks in the small-world regime. This topology balances short average path lengths (via shortcuts that connect otherwise distant agents) with high clustering (which supports local reinforcement). By contrast, scale-free networks do not support the same transition dynamics. The key difference lies in degree heterogeneity. Our model assumes random seeding of initial adopters: under strong heterogeneity, outcomes depend heavily on whether seeds fall on hubs or peripheral nodes. If seeding happens to include a hub, diffusion proceeds readily; if, as is likely, seeds lie on low-degree nodes, diffusion stalls unless conformism is very low or convenience is very high (Fig. \ref{fig:scalefree}, \ref{fig:cm_3000}). Homogeneous connectivity in Watts–Strogatz networks makes random seeding less sensitive to the precise location of initial adopters, thereby supporting the transition.

In our model, contagion regimes emerge endogenously rather than being pre-specified \cite{complexcontagion}. When the new norm is highly convenient and conformism is low, agents require little reinforcement, and increasing weak ties accelerates spread, consistent with simple contagion. When the convenience gap between the new norm A and the old norm B narrows, increasing the rewiring probability alone is insufficient (see Fig. with \(\epsilon = 0.2\)). Time to adoption decreases up to a threshold in \(p\) and then increases, indicating a trade-off between global reach via shortcuts and local reinforcement via clustering. When utilities are close, adoption requires more reinforcement from neighbors, and diffusion exhibits complex-contagion-like signatures—again without any explicit adoption threshold encoded in the model.

We also show that salient features of norm dynamics observed empirically can be reproduced by a parsimonious model. Prior work reproduces tipping points \cite{muthukrishna2020} and complex contagion \cite{horsevad2022transition} by specifying explicit thresholds or reinforcement rules at the micro level. Here, different contagion-like regimes arise endogenously without such rules. A relatively simple game-theoretic model on an abstract complex topology—with two dynamical parameters and one topological parameter—suffices to recover realistic macroscopic patterns.
The parsimony of our model is a strength, but it entails structural and behavioral simplifications. Moreover, we employed static networks, whereas real societies evolve. Testing the model on adaptive or temporal networks would yield a more faithful representation and enable robustness checks.
We also assigned conformity and convenience homogeneously at the network level. In reality, individuals differ in values and behavior: conformism and perceived convenience should be modeled as agent-level heterogeneous attributes.

Finally, the emergence of simple- and complex-contagion-like dynamics should be interpreted cautiously. Our macroscopic diffusion patterns are consistent with these mechanisms, but they do not imply that identical microscopic processes occur in real societies. Indeed, a key result is that such patterns can arise without explicit individual-level contagion rules.

In sum, we isolate a small set of features that enable study of the mechanisms underlying emergent behavior. Future work can attempt to test how this model behaves under more realistic circumstances to check if this dynamic survives when we relax these assumptions.

\section{Conclusions}

The aim of this paper is to identify the conditions under which new norms emerge in structured populations. We focus on three dimensions—network topology, conformism, and convenience—to analyze their contributions to emergence. We also examine whether simple- and complex-contagion regimes can arise endogenously from their interactions.

First, model outcomes are strongly influenced by network structure: topology is not merely a background for the dynamics but a fundamental determinant of whether dynamics can emerge. Within the model space we test, the small-world regime of the Watts–Strogatz network provides structural conditions that support transition dynamics. Given an appropriate topology, systematically varying norm convenience and conformism reveals a transition curve in parameter space separating persistence of the \textit{status quo} from the emergence of a new norm. Finally, distinct diffusion regimes emerge within the same model depending on the region of parameter space: simple- and complex-contagion dynamics arise as system properties rather than as pre-specified rules, emerging without any explicit adoption threshold encoded ex ante.

In sum, a simple model with a few chosen parameters generates diverse collective behaviors. Moreover, the diffusion regime need not be imposed \textit{ex ante}; it can emerge endogenously from social structure, the individual advantages of change, and social influence.

\section*{Appendix}

The final master equation \eqref{eq:final} can be easily obtained by Eq. \eqref{eq: me} just by considering the explicit expressions for the transition rates for the VM and FR dynamics. The VM dynamics is purely imitative, so the mean field probability that the number of the agents adopting the $A$ strategy increases/decreases by 1 unit is given directly by the relative size of the two populations: in order to go from $n_A$ to $n_A+1$ we have to pick one agent (over $N$) adopting the $A$ strategy, then an agent adopting the $B$ strategy (over $N-1$), and viceversa for the inverse transition. Therefore, since it is $n_B=N-n_A$, the transition rates can be written down as

\begin{equation}
T^{\pm}_{VM}(n_A\pm1|n_A)=\frac{n_A}{N}\frac{N-n_A}{N-1}=\frac{N-n_A}{N}\frac{n_A}{N-1}
    \label{App1}
\end{equation}

\noindent (notice how the transition rates to increase or decrease $n_A$ are actually equal). Analogously, the transition rates for the number of agents adopting the $A$ strategy going from $n_A\pm1$ to $n_A$ are

\begin{equation}
T^{\pm}_{VM}(n_A\pm1|n_A)=\frac{n_A\pm1}{N}\frac{N-n_A\mp1}{N-1}=\frac{N-n_A\mp1}{N}\frac{n_A\pm1}{N-1} \ . 
    \label{App2}
\end{equation}

\noindent Concerning the strategic transitions given by the Fermi rule, with the mean field approximations they are just the same VM transitions times the Fermi probability $[1+\exp{(-\beta\Delta\pi)}]^{-1}$, where $\Delta\pi$ is the average difference between the payoffs of the two strategies given the initial and final states. For instance, the FR rate for the transition $n_A\rightarrow n_a+1$ will be

\begin{equation}
T^+_{FR}(n_A+1|n_A) = \frac{T^+_{VM}(n_A+1|n_A)}{\exp{(-\beta\Delta\pi(n_a\rightarrow n_A+1))}} \ , 
    \label{App3}
\end{equation}

\noindent where it is easy to get that $\Delta\pi(n_a\rightarrow n_A+1)=(1+\epsilon)n_A-(N-n_A)$.

Finally, Eq. \eqref{eq:final} is obtained inserting these transition rates in Eq. \eqref{eq:Tmeno} and doing the thermodynamical limit $N\rightarrow+\infty$, where $n_A/N=\rho$, $N\pm1\simeq N$, and also $n_A\pm1\simeq n_A$: this last approximation makes the final equation valid only in the open interval $\rho\in(0,1)$.

\

\section*{Acknowledgements}
C.L. would like to thank Prof. Manlio De Domenico for his valuable feedback, insightful suggestions, and constructive discussions during the development of her original thesis, as well as for his encouragement in pursuing further research at the CNR.

\

\bibliographystyle{plain} 
\bibliography{refs} 

\end{document}